\documentclass[aps,prb,showpacs,twocolumn]{revtex4}

\usepackage{amsmath}

\begin{document}

\input psfig.sty

\title{Breakup of a Stoner model for the 2D
ferromagnetic quantum critical point}
\author{M.\ Dzero
\footnote{Present address: U.S. Department of Energy Ames
Laboratory, Ames, IA 50011. E-mail: dzero@ameslab.gov.}$^{,1}$ and
L.\ P.\ Gor'kov$^{1,2}$} \affiliation{$^1$National High Magnetic
Field Laboratory, Florida State University, Tallahassee, FL 32310,
USA} \affiliation{$^2$L.D. Landau Institute for Theoretical
Physics, Russian Academy of Sciences, 117334 Moscow, Russia}

\begin{abstract}
Re-interpretation of the results by [A. V. Chubukov \emph{et.
al.}, Phys. Rev. Lett. {\bf 90}, 077002 (2003)] leads to the
conclusion that ferromagnetic quantum critical point (FQCP) cannot
be described by a Stoner model because of a strong interplay
between the paramagnetic fluctuations and the Cooper channel, at
least in two dimensions.
\end{abstract}
\maketitle

Recent experimental observations of superconductivity in a close
proximity to the ferromagnetic quantum critical point (QCP) in a
few itinerant ferromagnets such as heavy fermion compounds UGe$_2$
\cite{Saxena}, URhGe \cite{Aoki} and also in ZrZn$_2$
\cite{Pfleiderer} have revived an interest to whether the magnetic
fluctuations in the vicinity of ferromagnetic QCP provide a
fundamental mechanism for the "$p$-wave" superconductivity. The
idea of the exchange by paramagnons turned out to be crucial for
understanding the mechanism underlying the superfluid state in
$^3$He at low temperatures \cite{Leggett}. In a broader context,
relevance of paramagnetic fluctuations to the appearance of
superconductivity at the border of a magnetic transition has been
discussed in \cite{Mathur} and \cite{Joerg} where reader will find
the history of the problem and the comprehensive list of
references.

Starting already with the one of the first papers on the subject
\cite{Appel} there were numerous efforts to evaluate the
superconducting transition temperature mediated by strong
paramagnetic fluctuations for both ferromagnetic and paramagnetic
phases \cite{Bedell, Millis, Monthoux}. Most of these results bear
the numerical character, and this obscures the fact that there is,
as we believe, some flaw in the very model. Most of the authors
deal with the Stoner model where an instability of the
paramagnetic itinerant state comes about with the increase of the
on-site Hubbard interaction, $U$, that leads to the appearance of
ferromagnetism. Variation in the value of the Stoner factor
governs then changes in a value of the Curie temperature by
varying an external parameter (in a more general form these ideas
can be formulated in terms of the Fermi liquid theory
\cite{Kondratenko} by involving the Pomeranchuk instability
\cite{Pomeranchuk, Abrikosov}). Below we will try to demonstrate
that such a model is not self-consistent, at least for the $2D$
systems.

The first question one faces while discussing any phase transition
is how close one can approach the line of a transition. The
problem of a singularity, or magnitude of fluctuations, looks
simpler near the ferromagnetic QCP (the ending point of the phase
diagram at $T=0$). The outcome of the analysis done by Hertz
\cite{Hertz} and recently by Millis \cite{Millis2} is that near
such QCP fluctuations retain their mean field character due to an
increase in effective dimensionality to account for an involvement
of the frequency variable at zero temperature. We argue, that the
hypothesis of the ferromagnetic QCP with the paramagnon propagator
\cite{Hertz} would lead to the developing of superconducting
fluctuations at such a scale which breaks the validity of Hertz
analysis far away from the vicinity of the imaginable QCP. In
other words, a Stoner like ferromagnetic QCP is not
self-consistent \emph{namely because it seemingly leads to such
strong pairing fluctuations} that make incorrect independent
analysis of the spin "zero sound" and Cooper channels. Below, we
would like to prove our point by re-interpreting the results of
the recent paper by Chubukov \emph{et al.}~\cite{Chubukov}.

The emphasis in Ref. \cite{Chubukov} has been put to demonstrate
that superconducting transition near the ferromagnetic QCP may
turn out to be of the first-order. The authors of \cite{Chubukov}
used the standard ansatz for the longitudinal fluctuations
propagator \cite{Hertz}. To get rid of the so-called non-adiabatic
corrections and to reduce the gap equations to the well known form
of the strong coupling ones ("Eliashberg" equations), the authors
made an assumption that the interaction of electrons with the spin
fluctuations is weak. A minor change below makes it more
convenient to overview the physical picture of their model as a
whole without resorting to numerical calculations.

Let us introduce the exchange part of interaction between the two
electrons, $\widehat{I}({\mbox{\boldmath $r_1 - r_2$}})$ as:
\begin{equation}
\widehat{I}({\mbox{\boldmath $r_1 - r_2$}}) = - I({\mbox{\boldmath
$r_1-r_2$}})\left(\mbox{\boldmath$\widehat{\sigma}_1\cdot\widehat{\sigma}_2$}\right)
\label{one}
\end{equation}
and assume that $\hat{I}({\mbox{\boldmath $r_1 - r_2$}})$ has the
ferromagnetic sign and bears a \emph{long-range} character. After
summing up all diagrams in the zero-sound channel the same way it
was done in \cite{Hertz}, one arrives to the following
longitudinal spin-spin fluctuation propagator:
\begin{equation}
\widetilde{I}(\mbox{\boldmath $q$}; \omega_n) =
\frac{I(\mbox{\boldmath
$q$})}{1-2~\nu\left(\varepsilon_F\right)I(\mbox{\boldmath
$q$})\left[1 - a\frac{|\omega_n|}{v_F q}\right]}, \label{two}
\end{equation}
where $I(\mbox{\boldmath $q$})$ is the Fourier component for the
interaction (\ref{one}), $w_n=\pi{T}(2n+1)$ is the Matsubara
frequency, $\nu(\varepsilon_F)$ is the density of states at the
Fermi level and the value of $a$ depends on the dimensionality of
the system: $a=\pi/2$ for 3D and $a=1$ for 2D.

Cancellation in the denominator of (\ref{two}) at $q\to{0}$ and
$\omega_n/v_Fq\ll{1}$ leads to the Stoner criterion:
$1-2~\nu(\varepsilon_F)I(0)\to{0}$. The factor in the square
brackets in denominator in (\ref{two}) is nothing but the electron
polarization operator $\Pi(\mbox{\boldmath $q$};\omega_n)$:
\begin{equation}
\Pi(\mbox{\boldmath
$q$};\omega_n)=\frac{T}{(2\pi)^d}\sum\limits_{\Omega_l}^{}
\int{d^d\mbox{\boldmath $l$}}~G(\mbox{\boldmath
$q+l$};\omega_n+\Omega_l)\cdot{G}(\mbox{\boldmath $l$};\Omega_l).
\tag{2'} \label{twoprime}
\end{equation}
proportional to the generalized electron spin susceptibility and
calculated at small enough $q$ in the limit
$|\omega_n|\ll{v_F{q}}$ (at this relation $\omega_n$ and $q$
appear in all equations below). In the opposite limit
$\Pi(\mbox{\boldmath $q$};\omega_n)$ decreases to zero, and the
Stoner like enhancement rapidly disappears at
$\omega_n{\sim}v_F{q}$.

According to \cite{Hertz}, Eq. (\ref{two}) for the ferromagnetic
fluctuations does not experience renormalization at $T=0$ and
$1-2~\nu(\varepsilon_F)I(0)\to{0}$. To agree the form of Eq.
(\ref{two}) with the similar expressions in \cite{Chubukov} we
assume that $I(q)$ rapidly decreases with an increase in $q$. To
be more specific, we accept the following notation:
\begin{equation}
2I(q)\nu(\varepsilon_F)\equiv{I_0}
\exp\left[-q^2/(p_F\vartheta)^2\right], \label{three}
\end{equation}
where
\begin{equation}
\vartheta\equiv\left(R{p_F}\right)^{-1}\ll{1} \label{four}
\end{equation}
is the parameter sharing the long range character of (\ref{one}),
which in turn leads the small angle scattering to prevail in Eq.
(\ref{two}). The interaction (\ref{one}) is isotropic in the spin
space. It is straightforward to account in Eqs.
(\ref{one},\ref{two}) for the presence of the magnetic anisotropy
\cite{Millis}.

As it was first pointed out in \cite{Schrieffer}, spin
fluctuations produce two effects in case of the phonon-mediated
($s$ - wave) pairing: they add to renormalization of the electron
self-energy and they provide the pair breaking mechanism for the
$S=0$ Cooper pair. Pair breaking effects are basically the same
even for a triplet ($S=1$) pairing, however, the exchange between
two electrons by longitudinal paramagnetic fluctuations leads to
the attractive interaction in the triplet $(S=1)$ channel. For the
electron's Green function in the paramagnetic phase,
$G^{-1}(\mbox{\boldmath $p$};\omega_n) = i\omega_n -
\varepsilon(\mbox{\boldmath $p$})-\Sigma(\mbox{\boldmath
$p$};\omega_n)$, we write:
\begin{widetext}
\begin{equation}
\Sigma(\mbox{\boldmath $p$};\omega_n) = \frac{c}{(2\pi)^d}
{T}\sum\limits_{\omega_{n'}}^{}\int{d^d\mbox{\boldmath $p_1$}}
\widetilde{I}(\mbox{\boldmath
$p-p_1$};\omega_n-\omega_{n'})G(\mbox{\boldmath
$p_1$};\omega_{n'}), \label{five}
\end{equation}
where $d$ is the dimensionality of the problem, and the
coefficient $c$ in (\ref{five}) depends on whether the exchange is
isotropic ($c=3$) or a strong magnetic anisotropy is present
($c=2$ for "easy plane" and $c=1$ for "easy axis"). For the sake
of simplicity we take $c=1$ ("easy axis") in order to avoid
additional complications related to the possibility of a
first-order superconducting transition.

The superconducting order parameter
$\widehat\Delta_{\alpha\beta}(\mbox{\boldmath $p$};\omega_{n})$ is
chosen below in the form:
\begin{equation}
\widehat\Delta_{\alpha\beta}(\mbox{\boldmath $p$};\omega_{n})=
i\left[(\mbox{\boldmath
$\widehat{\sigma}\cdot{d}$}(\mbox{\boldmath
$p$};\omega_{n}))\widehat{\sigma}_y\right]_{\alpha\beta}
\label{six}
\end{equation}
and near transition the vector $\mbox{\boldmath $d(p$};\omega_n)$
satisfies the linear equation:
\begin{equation}
\mbox{\boldmath
$d(p$};\omega_n)=\frac{T}{(2\pi)^d}\int{d^d}\mbox{\boldmath
$p_1$}\widetilde{I} (\mbox{\boldmath
$p-p_1$};\omega_n-\omega_{n'})~G(\mbox{\boldmath
$p_1$};\omega_{n'})~G(-\mbox{\boldmath $p_1$};-\omega_{n'})
\mbox{\boldmath $d(p_1$};\omega_{n'})
\label{seven}
\end{equation}
\end{widetext}
which would serve to determine the dependence of the
superconducting transition temperature on the proximity to the QCP
in Eq. (\ref{two}). According to \cite{Chubukov}, the vicinity of
the ferromagnetic QCP where onset of superconductivity acquires a
non-perturbative character is much broader in 2D then in 3D.
Consequently, our discussion will be restricted to the 2D
situation only.

Let us express vector $\mbox{\boldmath $q$}$ in Eq. (\ref{three})
through the angle of scattering along the Fermi surface,
$\varphi\ll{1}$:
\begin{equation}
\mbox{\boldmath $q$}^2 = {p_F}^2\varphi^2 \label{eight}
\end{equation}
In Eqs. (\ref{five},\ref{seven}) one can integrate over the energy
variable component, $\xi_1=v_F(p_1-p_F)$. Dependence on
$\mid\mbox{\boldmath $p$}\mid$ in the order parameter
$\widehat\Delta_{\alpha\beta}(\mbox{\boldmath $p$};\omega_{n})$
can be neglected because of smallness of the scattering angle,
$\varphi$:
\begin{equation}
\mbox{\boldmath $d$}(\mbox{\boldmath $p$};\omega_n)\Rightarrow
\mbox{\boldmath $d$}(\omega_n). \label{nine}
\end{equation}
Introducing the notation for proximity to the QCP
\begin{equation}
1-I_0=\tau\ll{1} \label{ten}
\end{equation}
after simple calculation Eq. (\ref{seven}) acquires the form:
\begin{equation}
\mbox{\boldmath $d$}(\omega_n)=
T_c\sum\limits_{\omega_{n'}=-\infty}^{\infty}
\frac{i\pi\langle\widetilde{I}\rangle_{\omega_n-\omega_{n'}}}
{i\omega_{n'}-\Sigma(\omega_{n'})} \mbox{\boldmath
$d$}(\omega_{n'}), \label{eleven}
\end{equation}
where
\begin{equation}
\langle\widetilde{I}\rangle_\omega=\frac{\vartheta}{2\pi}
\int\limits_{0}^{\infty}\frac{ds}{\tau + s^2 +
\frac{\omega}{v_Fp_F}\frac{1}{s\vartheta}} \label{twelve}
\end{equation}
comes about after integrating (\ref{two}) over $\varphi$ and
making use of Eqs. (\ref{three},\ref{four},\ref{eight}) in the
expansion:
\begin{equation}
1-2I(\mbox{\boldmath $q$})\nu(\varepsilon_F)\simeq\tau +
\left({\varphi}/{\vartheta}\right)^2 \label{thirteen}
\end{equation}

In equations above we have neglected all the vertex corrections to
the bare vertices. With the help of the explicit expressions for
$\widetilde{I}(\mbox{\boldmath $p$};\omega_n)$, it will be easy to
verify that the higher order corrections are small provided that
our parameter $\vartheta$ above is small.
\begin{figure}[h]
\vspace{-0.25cm}
\centerline{\psfig{file=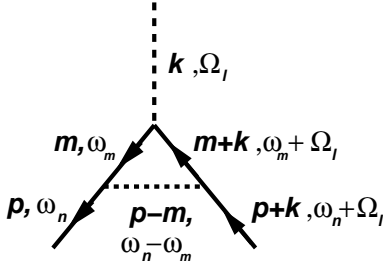,height=3.5cm,width=5cm}}
\caption{Vertex correction to the interaction. Dashed and solid
lines are the paramagnon and fermionic propagators,
respectively.}\label{vertex} \vspace{-0.25cm}
\end{figure}
Straightforward calculation of the first vertex correction to the
self-energy $\Sigma(\mbox{\boldmath $p$};\omega_n)$ (\ref{five})
shown in Fig. \ref{vertex}
\begin{eqnarray}
\delta\Gamma^{(1)}=\frac{i\pi{T}}{2}\sum\limits_{\omega_m}^{}
\left[sign(\omega_m+\Omega_l)-sign(\omega_m)\right]
\nonumber\times\\
\int\limits_{0}^{\infty}\frac{d\varphi}{2\pi}\frac{1}{(\mbox{\boldmath
$v$}_F\mbox{\boldmath $k$} -
i\Omega_l)(\tau+\varphi^2+\frac{\mid{\omega_m-\omega_n\mid}}{2\varepsilon_F\vartheta\varphi})}
\nonumber
\end{eqnarray}
gives the following order of magnitude estimate:
\begin{equation}
\delta\Gamma^{(1)}\sim\left({\vartheta}/{\sqrt{\tau}}\right)\cdot\left({\Omega_l}/{v_Fk}\right)
\tag{13'} \label{thirteenprime}
\end{equation}

In notations (\ref{twelve}) denominator in (\ref{eleven}) is:
\begin{equation}
i\omega_n-\Sigma(\omega_n) = i\omega_n +
i\pi{T}\sum\limits_{\omega_{n'}=-\omega_n}^{\omega_n}
\langle\widetilde{I}\rangle_{\omega_{n'}} \label{fourteen}
\end{equation}
Substitution $s=z\sqrt{\tau}$ transforms (\ref{twelve}) into:
\begin{equation}
\langle\widetilde{I}\rangle_{\omega}=\left(\frac{\vartheta}{2\pi\sqrt{\tau}}\right)
\int\limits_{0}^{\infty}\frac{zdz}{z(1+z^2) +
\frac{\omega}{\omega_0}\left(\frac{\vartheta}{4\sqrt{\tau}}\right)^3}
\label{fifteen}
\end{equation}
with
\begin{equation}
\omega_0=\varepsilon_F\vartheta^4/32. \label{sixteen}
\end{equation}
Frequency dependence in (\ref{twelve}) serves as the cut-off:
\begin{equation}
\langle\widetilde{I}\rangle_\omega\approx{1}/{3\sqrt{3}}\left({\omega_0}/{\omega}\right)^{1/3}
\label{seventeen}
\end{equation}
for
\begin{equation}
\omega>\omega_0\left({4\sqrt{\tau}}/{\vartheta}\right)^3\simeq{2}\varepsilon_F\vartheta\tau^{3/2}
\tag{17'} \label{seventeenprime}
\end{equation}
\begin{figure}[h]
\vspace{-0.5cm}
\centerline{\psfig{file=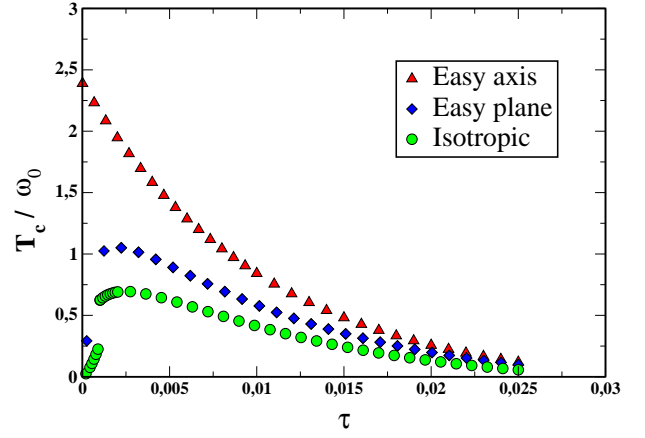,height=7cm,width=9cm,angle=-90}}
\caption{Numerical solution of Eq. (\ref{eleven}) for $T_c$ as a
function of $\tau$ for the various values of magnetic anisotropy
constant, $c$ (see inset). Maximum in $T_c$ for easy plane and
isotropic cases corresponds to the onset of the first-order
transition into superconducting state.}\label{fig1}
\vspace{-0.75cm}
\end{figure}

Let us consider the paramagnetic phase ($\tau{>0}$) far enough
from the QCP. Renormalization of the self-energy part
$\Sigma(\omega_n)$ is small, so that Eq. (\ref{eleven}) converts
into the weak-coupling problem:
\begin{equation}
1=\left(\frac{\theta}{4\sqrt{\tau}}\right)ln\left[\frac{\widetilde{\omega}}{T_c}\right],
\label{eighteen}
\end{equation}
where $\widetilde{\omega}$ is of the order of
$\omega_0\left(\frac{4\sqrt{\tau}}{\vartheta}\right)^3$ (see Eq.
(\ref{seventeenprime})). When $\sqrt{\tau}$ is further decreased,
so it becomes comparable to $\vartheta$, the strong-coupling
regime sets in with the cut-off frequency
$\widetilde{\omega}\simeq\omega_0$ in (\ref{eighteen}). This leads
to the new and important changes in $\Sigma(\omega)$. As it has
already been discussed in \cite{Chubukov}, one has the following
asymptotic behavior of the self-energy part:
\begin{equation}
\Sigma(\omega)=\left\{
\begin{array} {r@{\quad}}
\omega\left({\vartheta}/{(4\sqrt{\tau})}\right), ~~\omega\ll{2}\varepsilon_F\vartheta\tau^{3/2}, \\
\left(\frac{3\sqrt{3}}{4}\varepsilon_F\vartheta^4\right)^{1/3}\cdot\omega^{2/3},
~~\omega\gg{2}\varepsilon_F\vartheta\tau^{3/2}
\end{array}
\right. \label{nineteen}
\end{equation}
The second asymptotic would signify appearance of the non-Fermi
liquid regime in the close proximity to the QCP. It is seen, for
fixed $\vartheta$, that the overall behavior of the critical
temperature is defined by the value of the "coupling" constant,
$\lambda$:
\begin{equation}
\lambda = {\vartheta}/{4\sqrt{\tau}} \label{twenty}
\end{equation}
As it is readily seen from Eq. (\ref{eighteen}), $T_c\sim\omega_0$
at $\lambda\sim{1}$.

Equation (\ref{eleven}) for $T_c$ as a function of $\tau$ was
solved numerically with the solution shown on Fig. \ref{fig1}. At
$\tau\equiv{0}$ we obtain:
\begin{equation}
T_c\simeq{2.5\omega_0}\simeq{0.08\varepsilon_F\vartheta^4}.
\label{twentyone}
\end{equation}
Thus we conclude that $T_c$ is finite at $\tau=0$ and reaches the
energy scale of the order of $\omega_0$ already at
$\lambda\sim{1}$. The latter is also true for other cases shown on
Fig. 2.

Let us therefore keep $\lambda\sim{1}$ and start gradually
increasing value of the model parameter $\vartheta$. At
$\vartheta\to{1}$ we should return back to the Stoner model with
ordinary local interaction which is discussed in \cite{Hertz}. We
see from (\ref{thirteenprime}), that non-adiabatic corrections
remain of the order of one at $\lambda\sim{1}, \vartheta\sim{1}$
and, hence, one expects that their exact treatment would not
change qualitatively the estimate $T_c\sim\varepsilon_F$ from Eq.
(\ref{twentyone}). On the other hand, at $\lambda\sim{1}$ and
$\vartheta\to{1}$, $\tau=\vartheta^2/16\lambda^2\sim{1}$, and one
cannot come close to QCP without forming a superconducting ground
state, which in turn would change the polarization operator
(\ref{twoprime}). Obtaining such a high values for $T_c$
energetically so far away from the originally accepted position of
the QCP ($\tau=0$) shows the intrinsic contradiction of the local
Stoner model \cite{Hertz}.
At such a high energy scale the assumed proximity to a "QCP" seems
to be irrelevant for physical properties of the system.
Mathematically, large $T_c\sim{\varepsilon_F}$ or/and large
$\Delta$ lead to the change in the polarization operator Eq.
(\ref{twoprime}) to account for an interplay between the
zero-sound and superconducting channels. The polarization operator
(\ref{twoprime}) is modified due to the presence of anomalous
Gor'kov functions $F, F^{\dagger}$ in superconducting state at
$T=0$. For $\vartheta\sim{1}$ and
$\Delta\sim{T_c}\sim\varepsilon_F$ this introduces such a change
in polarization operator that significantly reduces cancellation
in the Stoner factor.

Remember now that at $\omega=0$ and $v_Fq\to{0}$ expression
(\ref{twoprime}) used to describe a behavior of magnetic
susceptibility near a ferromagnetic QCP. Similarly, modified
polarization operator is proportional to electronic susceptibility
in \emph{superconducting state}. The latter would not go to zero
at $T=0$ as it does for the $s$-wave pairing, even though it does
not equal to normal susceptibility neither in any triplet state.
The Stoner cancellation does not occur. Therefore, once
superconductivity (\ref{eighteen}) arises at $\tau > 0$ in the
framework of the model with $\vartheta\ll{1}$, at $T=0$ the ground
superconducting state as a function of external parameter
continues to be stable while entering into the ferromagnetic state
$(\tau < 0)$ well beyond "QCP". This suggests a first order like
phase competition between superconductivity and ferromagnetism at
low temperatures.

All that has been said above poses a few questions. First,
numerical calculations (see Ref. \cite{Bedell,Millis,Monthoux})
for $T_c$ making use of an exchange by longitudinal spin
fluctuations give rather low values of critical temperature
compared to the values of the bandwidth or the Fermi energy
without special assumption of small angle scattering (the authors
have been solving basically the same equations, i.e. no vertex
corrections have been included). We believe, this is a result of
some numerical smallness, such as the factor $1/32$ in Eq.
(\ref{sixteen}). This smallness may restore QCP within some
vicinity. Indeed, the low values of $T_c$ has been experimentally
observed in a number of systems among which are Sr$_2$RuO$_4$,
URhGe or $^3$He. In addition, in 3D the strong coupling regime of
the above model sets on only in very close proximity to the QCP
\cite{Chubukov}. Note that the model itself does not determine the
spatial symmetry of the triplet Copper pair wave-function:
exchange by paramagnons is not the only interaction between the
electrons in the Cooper channel. Thus, the paramagnon contribution
in our model comes together with other interactions,
$\Gamma_l\sim{1}$:
\begin{equation}
(3D): ~-\vartheta^2|{ln(\tau)}| + \Gamma_l;
~~(2D):~-\frac{\vartheta}{\sqrt{\tau}} +
\Gamma_l,\label{twentythree}
\end{equation}
where $\Gamma_l$ may be positive (here $\Gamma_l$ is a proper
harmonics designed by the exact symmetry of the superconducting
pairing). Onset of attraction in (\ref{twentythree}) takes place
closer to $\tau=0$ and an effective interaction
(\ref{twentythree}) remains reasonably weak ($\sim{1}$) in its
vicinity ~\cite{NotaBene}.

To summarize, we have shown that competition with the
superconductivity channel makes the Stoner model for ferromagnetic
quantum critical point not self-consistent. FQCP can be realized
only due to the presence of a numerically small parameter or other
repulsive interactions in the triplet channel that weaken the
attraction mediated by paramagnons, at least in 2D.

The authors thank A. Chubukov, A. Finkel'stein and D. Morr for
fruitful discussions.

This work was supported by DARPA through the Naval Research
Laboratory Grant No. N00173-00-1-6005 (M.D.) and by the NHMFL
through the NSF cooperative agreement DMR-9527035 and the State of
Florida (L.P.G.).

\end{document}